\documentclass[aps,pra,twocolumn,showpacs,superscriptaddress,floatfix,10pt]{revtex4-1}
\usepackage{framed}
\usepackage{graphicx}
\usepackage{xfrac}
\usepackage{amsmath}
\usepackage{epsfig}
\usepackage{helvet}
\usepackage{amssymb}
\usepackage{framed}
\usepackage{hyperref}

\begin{document}

\title{Cutoff Dependence and Complexity of the CFT$_2$ Ground State}

\author{Bowen Chen}
\author{Bart{\l}omiej Czech}
\author{Zi-zhi Wang} 
\affiliation{Institute for Advanced Study, Tsinghua University, Beijing 100084, China}
\vskip 0.25cm

\begin{abstract}
\noindent
We present the vacuum of a two-dimensional conformal field theory (CFT$_2$) as a network of Wilson lines in $SL(2,\mathbb{R}) \times SL(2,\mathbb{R})$ Chern-Simons theory, which is conventionally used to study gravity in 3-dimensional anti-de Sitter space (AdS$_3$). The position and shape of the network encode the cutoff scale at which the ground state density operator is defined. A general argument suggests identifying the `density of complexity' of this network with the extrinsic curvature of the cutoff surface in AdS$_3$, which by the Gauss-Bonnet theorem agrees with the holographic Complexity = Volume proposal.
\end{abstract}


\maketitle

\textit{Introduction.---} Complexity has attracted much interest in holographic duality, with several proposals for how the concept is realized on the AdS \cite{volume, action, wdwvolume} and CFT \cite{robmichal, piopt} sides of the AdS/CFT correspondence. In some cases the two sides agree, but the match is not compelling because the CFT proposals involve many free parameters. We put forward a new approach to complexity in holographic CFTs, which is more rigid and more concomitant to the bulk than are currently studied proposals. 

The motivation for studying complexity comes primarily from time dependent processes such as the growth of black hole interiors \cite{volume, tomjuan}. However, if the core idea that complexity manifests itself holographically as spatial size is correct, it must also apply to static configurations. (In such contexts we speak of `complexity of formation,' see e.g. \cite{formation}.) We set out to study one such example: the pure three-dimensional anti-de Sitter spacetime (AdS$_3$) that is holographically dual to the ground state of a two-dimensional conformal field theory (CFT$_2$). Although pure AdS$_3$ is static and is not a black hole, it is a perfect arena for testing ideas about complexity because (i) it is the best understood bulk geometry that arises in holography, (ii) its properties are fixed by symmetry so the discussion does not depend on details of the CFT or the field content of the bulk theory, and (iii) many prior discussions of the complexity of formation studied AdS$_3$ as an example \cite{piopt, formation, example} so it is a standard test case. 

One consequence of conformal symmetry is that the CFT$_2$ vacuum `looks the same' over an infinite range of scales, so its complexity can only be infinite. For a finite measure of complexity, the ground state must be subject to an ultraviolet (UV) cutoff and the sub-cutoff data appropriately truncated. In holography, the choice of cutoff manifests itself as a large scale (infrared, IR) cutoff in the bulk; any gravity-side proposal for complexity must have a built-in dependence on the cutoff. But the meaning and implementation of the bulk cutoff have been extensively debated \cite{cutoffs} and it is at present unclear what notion of a bulk cutoff is most conducive for studying complexity.

\textit{The plan.}--- 
We first list some heuristics of what the cutoff should mean in the context of holographic complexity. A closely related point concerns the meaning of \emph{state} in \emph{state complexity}, on which we also make relevant remarks. We then present an object in $SL(2,\mathbb{R}) \times SL(2,\mathbb{R})$ Chern-Simons theory, whose properties match everything we expect from \emph{the CFT$_2$ ground state defined at a cutoff}. Symmetry and cutoff heuristics single out a unique candidate for quantifying the complexity of this object, which in bulk language ends up matching the spatial volume encircled by the cutoff surface. Our emphasis is not on the specifics of the CFT$_2$ ground state but on forging a novel approach to defining complexity, which departs from the currently dominant models.

\textit{The cutoff.}--- 
Consider a semi-classical, static, asymptotically AdS$_3$ geometry dual to some state of a holographic CFT$_2$. We stipulate the following about how an IR cutoff in the former relates to a UV cutoff in the latter:
\begin{enumerate}
\item In longitudinally homogeneous geometries radial curves manifest constant CFT cutoffs. 
\end{enumerate}
Examples include $u={\rm const.}$ lines in the Poincar{\'e} patch
\begin{equation}
ds^2 = (dx^+ dx^- + du^2) / u^2 \label{poincare}
\end{equation}
and $\chi = {\rm const.}$ curves in the BTZ / AdS$_3$-Rindler metric:
\begin{equation}
ds^2 = - \sinh^2\chi d\tau^2 + d\chi^2 + \cosh^2\chi dx^2.
\label{btzmetric}
\end{equation}
\begin{enumerate}
\setcounter{enumi}{1}
\item Coarsening the CFT UV cutoff pushes the bulk cutoff away from the boundary (deeper in the interior).
\item Maximally coarsening the CFT cutoff (so it encompasses the entire Cauchy slice) reduces the theory to s-waves (constant modes). This places the cutoff surface on the horizon if one exists.
\end{enumerate}
This follows as a limit of point 2 and from subregion duality \cite{subregion}. If \mbox{s-waves} defined a bulk cutoff shy of the horizon, how would we interpret the region between the horizon and the cutoff? Yet the cutoff cannot reach beyond the horizon lest it violate subregion duality. 
\begin{enumerate}
\setcounter{enumi}{3}
\item In locally AdS$_3$ geometries, if the bulk cutoff follows a geodesic $\gamma$ then the corresponding CFT cutoff scheme truncates the modes localized in the interval subtended by $\gamma$ to its s-wave sector.
\end{enumerate}
The last point follows from remark 3 after recognizing the geodesic $\gamma$ as the horizon of an AdS$_3$-Rindler geometry. The appropriate way of smearing local operators to produce s-waves on CFT intervals was discussed in \cite{stereoscopy}.

\textit{The state.}--- A density operator is a map, which sends operators to real numbers:
\begin{equation}
\rho\! :\,\, \mathcal{O} \to {\rm tr}(\rho\, \mathcal{O}) 
\label{densopdef}
\end{equation}
An emphasis on viewing states as functionals on operators, rather than simply members of the algebra of observables, is central to our thinking about complexity. 

In field theory, the state's dependence on the cutoff restricts the domain of map (\ref{densopdef}). Specializing to the holographic context, suppose the cutoff is presented as a smooth and convex curve in the bulk, parameterized by proper length $\lambda$. We do not assume that the cutoff is homogeneous, so the profile of the cutoff curve generically depends on space. We require the map $\rho$---the CFT$_2$ ground state at the given cutoff scale---to satisfy:
\begin{itemize}
\item[(i)] $\rho$ eats up multi-local operators $\mathcal{O}_1(\lambda_1) \ldots \mathcal{O}_n(\lambda_n)$ and returns $\langle 0 | \mathcal{O}_1(\lambda_1) \ldots \mathcal{O}_n(\lambda_n) | 0 \rangle$.
\item[(ii)] The arguments $\mathcal{O}_i(\lambda_i)$ of $\rho$ are labeled by their locations $\lambda_i$ on the bulk cutoff curve. 
\item[(iii)] If the curve follows a geodesic over some range of $\lambda$'s, $\rho$ takes at most one input $\mathcal{O}(\lambda)$ from that range.
\item[(iv)] Shifts in the bulk curve and the attendant transformations of $\rho$ enact local changes of scale.  
\end{itemize}
Requirement (iii) follows from point 4 in the cutoff discussion: more than one independent input from a geodesic segment would be outside the s-wave sector of the AdS$_3$-Rindler space lying beyond the geodesic.

We now show that 3-dimensional $SL(2,\mathbb{R}) \times SL(2,\mathbb{R})$ Chern-Simons theory contains an object, which automatically satisfies demands (i-iv). The Chern-Simons language is useful because it accommodates locally varying $SL(2,\mathbb{R}) \times SL(2,\mathbb{R})$ frames and CFT operators are sections of an $SL(2,\mathbb{R}) \times SL(2,\mathbb{R})$ bundle \cite{nakahara}---a technicality we explain below. 
Emphatically, we will make no reference to gravitational dynamics, for which the Chern-Simons formalism is conventionally deployed.

\textit{A review of $SL(2,\mathbb{R}) \times SL(2,\mathbb{R})$ gauge fields.---} 
We are interested in Lorentzian CFT$_2$s on a space with metric $ds^2 = dx^+dx^-$. Introduce an orthogonal semi-infinite axis $u$ and place the CFT at $u=0$. Let $A$ and $\bar{A}$ be two ${\bf sl}(2,\mathbb{R})$-valued connection one-forms in $(x^+, x^-, u)$-space, i.e. gauge fields whose components are linear combinations of $L_{-1}, L_0, L_1$. The latter generate the left- and right-moving global conformal symmetries and satisfy:
\begin{equation}
[L_n, L_m] = (n-m) L_{n+m}
\end{equation}
We choose a \emph{flat} configuration of $A$ and $\bar{A}$, which satisfies boundary conditions:
\begin{align}
\lim_{u \to 0} u A_+ = L_{-1} \quad & {\rm and} \quad \lim_{u \to 0} A_- = 0 
\label{abc} \\
\lim_{u \to 0} u \bar{A}_- = L_{1\phantom{-}} \quad & {\rm and} \quad \lim_{u \to 0} \bar{A}_+ = 0
\label{abarbc}
\end{align}
With this choice, Wilson lines (and their networks) compute CFT$_2$ ground state correlation functions \cite{earlier, jaredliam}.

\textit{The CFT meaning of Wilson lines.}--- In gauge theories, gauge-invariant quantities are supplied by path-ordered exponentials along paths $(x^+(\sigma), x^-(\sigma), u(\sigma))$: 
\begin{equation*}
P \exp\!\int_{\sigma_b}^{\sigma_e}\!\!\! d\sigma 
\left( \partial_\sigma x^+ A_+ \!+\!
\partial_\sigma x^- A_- \!+\! 
\partial_\sigma u\, A_u \right)
\times (A \leftrightarrow \bar{A})
\end{equation*}
Ordinarily the path must be closed and this is called a Wilson loop. In the present case the flatness of $A$ and $\bar{A}$ makes all Wilson loops trivial. When the path is open there is a residual gauge dependence at the endpoints, so arbitrary Wilson \emph{lines} are not gauge-invariant. The exception is when the endpoints are taken to the boundary where the gauge dependence is frozen by boundary conditions. Consequently, we focus below on Wilson lines with endpoints at the asymptotic boundary $u \to 0$.

Consider a path from $(x^+_b, x^-_b, u_0)$ to $(x^+_e, x^-_e, u_0)$. Because $A$ and $\bar{A}$ are flat, the exact course of the path is immaterial so we can pull it onto the $u=u_0$ plane. The Wilson line of $A$ evaluates to $\exp\!\big({L_{-1} (x^+_e - x^+_b)/u_0}\big)$, a finite translation by $(x^+_e - x^+_b)/u_0$. The $\bar{A}$-Wilson line performs a similar translation, only in $1/x^-$.
Indeed, any Wilson line is a finite $SL(2,\mathbb{R}) \times SL(2,\mathbb{R})$ transformation because it is a path-ordered exponential of an ${\bf sl}(2,\mathbb{R}) \times {\bf sl}(2,\mathbb{R})$ one-form. Due to the boundary conditions (\ref{abc}-\ref{abarbc}), Wilson lines that begin and end on the boundary are always finite translations in $x^+$ and in $1/x^-$.

We would like to compute expectation values of such Wilson lines in various $SL(2,\mathbb{R}) \times SL(2,\mathbb{R})$ representations. Representations of $SL(2,\mathbb{R})$ are classified by conformal dimensions $h$. Their multiplets are built up by acting with $L_{-1}$s on the highest weight state $|h\rangle \equiv \mathcal{O}_h(0) |0\rangle$. One example of an expectation value of our Wilson line is therefore:
\begin{equation}
\!\!\!\!\langle h | e^{L_{-1} (x^+_e - x^+_b)/u_0} | h \rangle 
\!=\! \langle 0 | \mathcal{O}_h(0) \mathcal{O}_h((x^+_e - x^+_b)/u_0) | 0 \rangle\,\,
\label{2ptcalc}
\end{equation}
We recognize that all CFT 2-point functions are expectation values of translations, taken over $SL(2,\mathbb{R})$ multiplets in various representations. Once again, equation~(\ref{abc}) ensures that boundary-pegged $A$-Wilson lines evaluate to the correct translations up to a rescaling by $u_0$, which we comment on below. For the $\bar{A}$-components of our Wilson lines, the conclusion is the same except the $SL(2,\mathbb{R})$ multiplets must be built in the $1/x^-$ conformal frame---by acting with $L_{+1}$s from the right on $\langle h| \equiv |h\rangle^\dagger$. 

\textit{The CFT meaning of networks.}--- 
For higher-point correlators, we combine Wilson lines to form networks---collections of lines joined by junctions. Such junctions were previously described in \cite{jaredliam}. In the discussion above, the flatness of $(A, \bar{A})$ allowed us to pull arbitrary Wilson lines onto $u={\rm const.}$ slices. We need to make sure that networks with junctions are similarly deformable. 

A junction merges two incoming Wilson lines into one outgoing line. Recall that an individual Wilson line carries an $SL(2,\mathbb{R}) \times SL(2,\mathbb{R})$ representation---a vector space populated by $|h, \bar{h} \rangle \equiv \mathcal{O}_{h,\bar{h}} |0\rangle$ and descendants. Two incoming Wilson lines carry two independent such multiplets, which a junction then sends to one multiplet:
\begin{align}
& \,\, \mathcal{O}_{h_1,\bar{h}_1}\!(v, \bar{v})\, |0\rangle_v 
\,\otimes\, 
\mathcal{O}_{h_2,\bar{h}_2}\!(w, \bar{w})\, |0\rangle_w \,\, 
\xrightarrow{~{\rm junction}~}
\label{junction} \\
& \int\! dz d\bar{z}\, \sum_{h,\bar{h}}
c^{h,\bar{h}}_{\phantom{h}\!h_1,\bar{h}_1;h_2, \bar{h}_2}\! (z,\bar{z},v,\bar{v},w,\bar{w})\,
\mathcal{O}_{h,\bar{h}}(z, \bar{z})\, |0\rangle_z
\nonumber
\end{align}
The notation $|0\rangle_{v}$ distinguishes the `ground state' on $v$-space from the `ground states' on $w$- and $z$-spaces---local copies of the CFT on which the $SL(2,\mathbb{R}) \times SL(2,\mathbb{R})$ multiplets live. Defining a junction boils down to choosing appropriate kernels $c^{h,\bar{h}}_{\phantom{h}\!h_1,\bar{h}_1;h_2, \bar{h}_2}\! (z,\bar{z},v,\bar{v},w,\bar{w})$ in (\ref{junction}). 

With an arbitrarily chosen kernel, networks of Wilson lines will not be deformable. A condition that guarantees deformability is this: if an $SL(2,\mathbb{R}) \times SL(2,\mathbb{R})$ transformation is simultaneously applied to the $v$- and $w$-space inputs in eq.~(\ref{junction}), its $z$-space output must come out transformed in the same way \cite{jaredliam}. An identical condition governs the Operator Product Expansion (OPE) of local operators in the CFT. Therefore, in a deformable junction, the dependence of $c^{h,\bar{h}}_{\phantom{h}\!h_1,\bar{h}_1;h_2, \bar{h}_2}\! (z,\bar{z},v,\bar{v},w,\bar{w})$ on its coordinate arguments is fixed by conformal kinematics up to overall constants $c^{h,\bar{h}}_{\phantom{h}\!h_1,\bar{h}_1;h_2, \bar{h}_2}$. We recognize that defining a deformable set of junctions leaves out the same flexibility as defining a CFT does: the undetermined data are OPE coefficients. 

Naturally, we choose Wilson line junctions to match the fusion algebra of the CFT. Thus, a network shown in the top left panel of the Figure represents a repeated application of the OPE expansion. Once again, this network can be pulled onto the boundary, in which case it \emph{is} the OPE expansion of the CFT. When the network is contracted at all endpoints with member states of various $SL(2,\mathbb{R}) \times SL(2,\mathbb{R})$ multiplets, it will by construction return the relevant multi-point correlation function.

\begin{figure}[!t]
\begin{center}
\includegraphics[width=1.0\columnwidth]{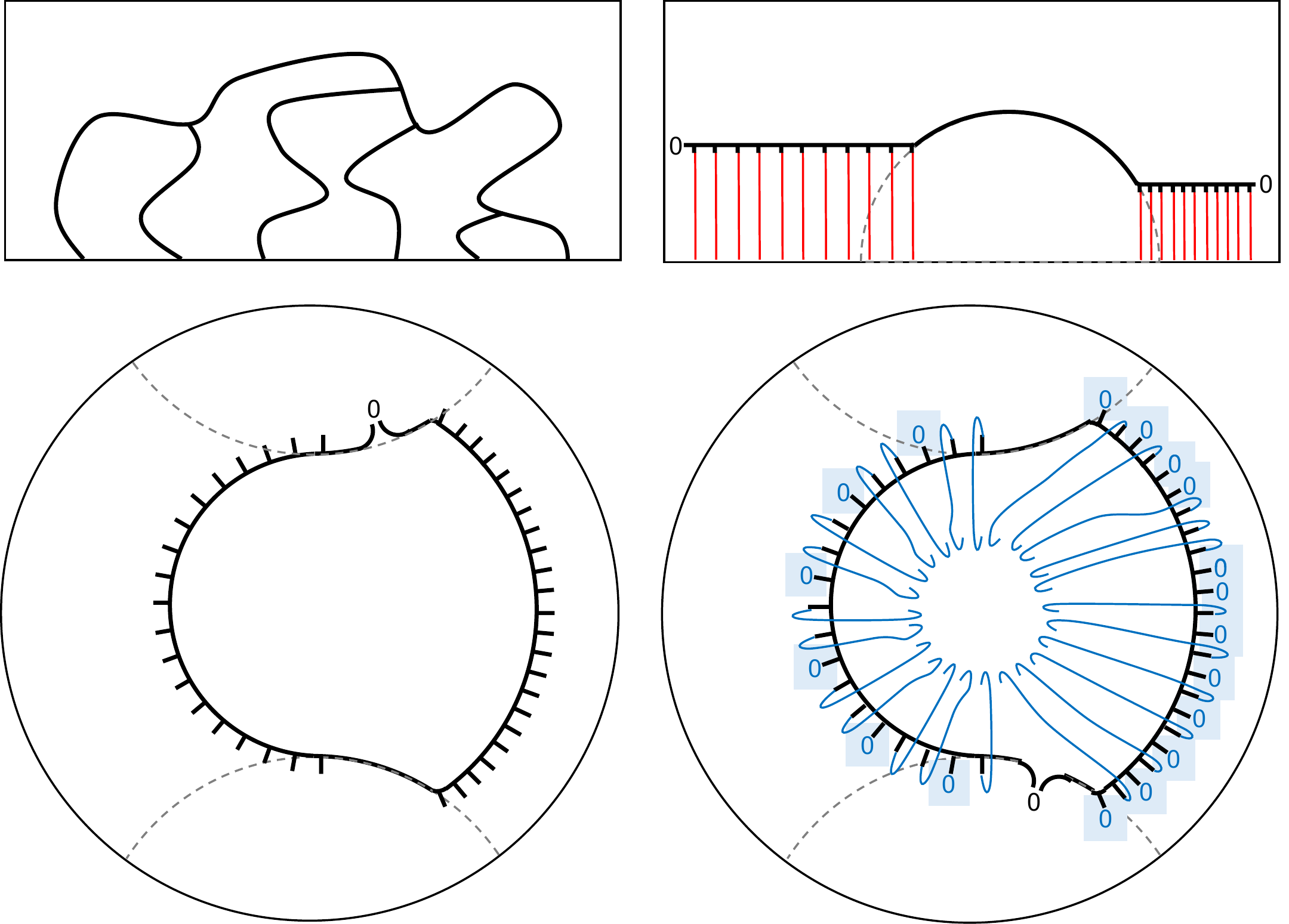}
\caption{All pictures are equal time snapshots of $(x^+,x^-,u)$-space. Top left: A network of Wilson lines, which computes CFT 6-point functions. Top right: A network which computes multi-point functions. After the red \emph{tentacles} are amputated, it becomes the ground state at the scale defined by the black line. The density of inputs on the network is $d\mathcal{C} \propto Kd\lambda$ so the network accepts no inputs on geodesic segments. Bottom: Two amputated networks in global AdS$_3$, which differ only by crossing symmetry, are shown in black. Crossing symmetry is implemented by shifting the feature ${\bf -}0\,0{\bf -}$, which implements the projector $|0\rangle \langle 0|$ in computing correlation functions. The blue features---Wilson lines and projections---convert the black network into one at a coarser, uniform cutoff. Projections are necessary to accord with $d\mathcal{C} \propto Kd\lambda$ because coarse-graining eliminates degrees of freedom. The coarse-graining scheme is highly non-unique; it varies with gauge choice for $(A, \bar{A})$ and with crossing symmetry. 
}
\end{center}
\end{figure}

\textit{The Chern-Simons field encodes the AdS$_3$ geometry.}---
The $SL(2,\mathbb{R}) \times SL(2,\mathbb{R})$ connection we exploit is most famous for a reformulation of pure gravity in asymptotically AdS$_3$ spacetimes \cite{witten}. When a flat $(A, \bar{A})$ which satisfies boundary conditions (\ref{abc}-\ref{abarbc}) is substituted into
\begin{equation}
g_{\mu\nu} = (\sfrac{1}{2})\, {\rm tr}_{\rm fundamental} (A-\bar{A})_\mu (A-\bar{A})_\nu \,,
\label{defmetric}
\end{equation}
it returns the metric of the Poincar{\'e} patch of AdS$_3$. The non-vanishing traces entering (\ref{defmetric}) are ${\rm tr}\, L_0^2 = 1/2$ and ${\rm tr}\, L_{-1} L_1 = -1$. Other locally AdS$_3$ geometries, including global AdS$_3$, arise from (\ref{defmetric}) with modified boundary conditions \cite{earlier, banados}. (The assertion that networks of Wilson lines compute CFT correlators still applies, with expectation values now taken in the vacuum on a cylinder or in Virasoro descendants.) As an example, the choice
\begin{align}
A & = (L_{-1}/u) dx^+ + L_0 d(\log u)
\label {achoice} \\
\bar{A} & = (L_{1}/u) dx^- - L_0 d(\log u)
\label{abarchoice}
\end{align}
produces metric~(\ref{poincare}) while other, gauge-equivalent choices produce the same geometry in different coordinates. The flatness of $(A, \bar{A})$, which expresses vacuum Einstein's equations with $\Lambda = -1$, follows from the classical equations of motion of Chern-Simons theory. 

Our discussion of CFT correlation functions made no reference to the quantum Chern-Simons theory and $(A, \bar{A})$ were only used as auxiliary technical tools. But because the AdS$_3$ spacetime dual to the ground state of any holographic CFT$_2$ is described by metric~(\ref{defmetric}), we may explain the meaning of bulk cutoff curves with reference to classical solutions such as (\ref{achoice}-\ref{abarchoice}). 

\textit{The CFT vacuum at a holographically specified scale.}--- Consider the network of Wilson lines pictured in the top right panel of the Figure. Assume the upper, black part of the network follows the bulk cutoff curve. Now sever all the red tentacles of this network to produce the \emph{amputated network}. (Amputation creates stumps that transform in dual representations---objects that eat up incoming representations and return numbers.) We claim that this amputated network is the ground state at the scale specified by the cutoff surface. More examples, which represent the vacuum on a cylinder, are shown in the bottom panels of the Figure. (Our amputated network shares certain features with the tensor network of \cite{cirac}.)

The amputated network manifestly satisfies conditions (i-iii). It eats up $SL(2,\mathbb{R}) \times SL(2,\mathbb{R})$ representations (delivered by incoming Wilson lines) to return appropriate multi-point correlators. Incoming lines are sprinkled over the cutoff curve except---as marked in the Figure---on geodesic segments. To verify (iv), inspect one of the amputated tentacles, say in the $u$-direction in gauge (\ref{achoice}):
\begin{equation}
\exp{\int_{u_0}^{u_*} (du/u) L_0} = (u_* / u_0)^{L_0}\,.
\label{deltascale}
\end{equation}
We recognize that the job of the tentacles is to bring $SL(2,\mathbb{R}) \times SL(2,\mathbb{R})$ representations from some reference scale $u_0$ to the scale of the cutoff curve, $u_*$. If we shifted the cutoff curve from $u_*$ to a new scale $u_*'$, individual stumps would get rescaled by $(u_*/u_*')^{L_0}$ to absorb the changed scales of their inputs; this is just what requirement (iv) stipulates. (Under coarse-graining, some stumps are also projected out; see the bottom right panel of the Figure and text below. Such projections are routine in tensor networks that model the RG behavior of CFT states \cite{projnetworks}.) Equation~(\ref{deltascale}) explains why the endpoints of the Wilson line in (\ref{2ptcalc}) were put at some arbitrary $u_0$. Sending them to the true asymptotic boundary ($u_0 \to 0$) would have dilated the computation by an overall infinite factor familiar from AdS/CFT.

The amputated network is not gauge-invariant. Its gauge dependence, dual to the gauge dependence of tentacles like (\ref{deltascale}), is an inalienable trait of renormalization. There is no preordained way of renormalizing CFT fields; doing so with $(u_*/u_0)^{L_0}$ is an artifact of gauge (\ref{achoice}-\ref{abarchoice}) and, by extension, of metric~(\ref{poincare}). This ambiguity of holographic RG manifests the technicality we mentioned---that CFT operators are sections of $SL(2,\mathbb{R}) \times SL(2,\mathbb{R})$ bundles---and reflects the rule of thumb that boundary global symmetries are bulk gauge symmetries. 

Finally, we comment on the loose ends of the amputated network. They carry trivial representations, which institute $\langle 0|$ and $| 0 \rangle$ in the computation of $\langle 0 | \mathcal{O}_1\! \ldots\! \mathcal{O}_n | 0 \rangle$. Had we joined the ends, $| 0 \rangle \langle 0|$ would be replaced with a sum over representations and the network would compute torus correlators instead. Shifting the loose ends implements crossing symmetry by reordering applications of the fusion algebra. We illustrate this with the two black networks in the bottom panels of the Figure.

\textit{What is the complexity of the amputated network?}--- We should view the amputated network as an \emph{algorithm} for computing CFT correlation functions---an algorithm for evaluating $\rho$ in equation~(\ref{densopdef}). This way of thinking is not unfamiliar: any tensor network is an algorithm in the same sense. The holographic Complexity = Volume proposal \cite{volume} was motivated \cite{tomjuan} by imagining a space-filling tensor network; the rationale was that volume $\sim$ \#(tensors) counts the steps of the algorithm. In the amputated network the steps are applications of fusion, so it is logical to quantify the complexity of the network as the count of OPE fusions to be carried out. Equivalently, we will count the \emph{inputs} (amputated legs) of the algorithm. Because no algorithm runs faster than linearly in the inputs, we need no further proof that our algorithm is optimal. 

Does any principle dictate how many Wilson lines the network can absorb? Thus far, our considerations have strongly relied on the global symmetry of the CFT$_2$. Because AdS is dual to the ground states of \emph{all} holographic CFTs, a geometric representation of their complexities---if it exists---must be fixed by conformal symmetry. If so, the measure of complexity $d\mathcal{C}$ on our amputated network must be an $SL(2,\mathbb{R}) \times SL(2,\mathbb{R})$ invariant of the cutoff curve. Curves confined to a static slice of AdS$_3$ have only two local invariants: the proper length element $d\lambda$ and $K d\lambda$, where $K$ is the extrinsic curvature. (Higher powers of $K$ are local invariants on smooth curves, but become ill-defined if we discretize the curve into a sequence of geodesic segments.) The density of complexity on the cutoff curve must therefore be a linear combination of the two: ${d\mathcal{C}}/{d\lambda} = \#_0 + \#_1 K$. 
As our network takes no inputs on geodesic segments where $K=0$, $\#_0$ must vanish so $d\mathcal{C} \propto K d\lambda$. For a sanity check, note that all equal time $u = {\rm const.}$ lines in metric~(\ref{poincare}) have $K=1$ and $d\lambda = dx/u$, so as expected $d\mathcal{C}/dx \propto 1/u$ in that case. 

\textit{Comparison with other proposals.}--- Specializing again to cutoff curves confined to the static slice, the Gauss-Bonnet theorem relates the complexity of the amputated network to the volume enclosed by the cutoff surface: 
\begin{equation}
\mathcal{C} \propto \int K d\lambda = \int (-R) dV + 2\pi =  \int dV + 2\pi
\end{equation}
In the convention of eq.~(\ref{defmetric}), the Ricci scalar on the static slice is $R = -1$. We will not attempt to explain the additive Euler term, which does not scale with the cutoff.

Our argument seems to support the Complexity = Volume proposal \cite{volume}, but it is not compelling evidence. When we take the cutoff curve off the static slice, symmetry allows a new contribution ${d\mathcal{C}}/{d\lambda} \propto \#_1 K + \#_2 \tau$, where $\tau$ is the Lorentzian `torsion' of the curve. Whatever combination of extrinsic curvature and torsion might quantify complexity, it will not match the maximal volume inside the curve, which is a more erratic quantity. 

Instead, we emphasize conceptualizing complexity as counting steps of an algorithm which, as in equation~(\ref{densopdef}), sends operators to numbers. This assumption tacitly undergirds equating complexity with counting nodes in tensor networks. Yet the circuit model makes an extra assumption: that all intermediate steps of the algorithm must also be interpretable as states of the theory---all the way back to a \emph{reference state} at which the algorithm is initialized. We think this is too restrictive. Our amputated network provides an example: it has no reference state and, if you interrupt it at an intermediate stage without inserting the trivial representation, it computes an iterated OPE expansion instead of a density matrix. 

\textit{Toward other states.---} Can the amputated network be adapted to excited states? One possibility is an algorithm which transports and fuses modular frequency modes of cutoff-sized intervals instead of $SL(2,\mathbb{R}) \times SL(2,\mathbb{R})$ representations. (Transport of modular modes was sketched in \cite{berry}.) Because fusing local operators automatically produces vacuum modular modes of intervals \cite{stereoscopy}, this idea is consistent with the model in this paper. Meanwhile, in holography modular modes localize in the bulk \cite{stereoscopy, tomaitor} on geodesics and other special loci, so the guess maintains contact with holographic proposals. We plan to investigate this idea in CFT$_2$ states produced by heavy operators, which are dual to conical defects and microstates of BTZ black holes.

\textit{Acknowledgements.---}
We thank Luis Apolo, Alex Belin, Vijay Balasubramanian, Jan de Boer, Adam Brown, Alejandra Castro, Shira Chapman, Muxin Han, Micha{\l} Heller, Eliot Hijano, Lampros Lamprou, Henry Lin, Juan Maldacena, Alex Maloney, Alexandros Mousatov, Robert Myers, Sepehr Nezami, Xiaoliang Qi, Robert Raussendorf, Moshe Rozali, Sukhbinder Singh, Wei Song, Leonard Susskind, Erik Verlinde, Herman Verlinde, and Gabriel Wong for illuminating discussions. We thank the organizers of the Quantum Information and String Theory workshop held at YITP in Kyoto where key parts of this work were carried out. BCz also gives thanks for hospitality to: the organizers of the Amsterdam String Workshop, of the Quantum Fields and Strings Workshop at Seoul National University, and of the Gravitational Holography program held at KITP Santa Barbara, as well as Stanford University, University of Science and Technology of China, South China Normal University and Perimeter Institute. BCz acknowledges support from the Visiting Fellowship of Perimeter Institute; research at Perimeter Institute is supported in part by the Government of Canada through the Department of Innovation, Science and Economic Development Canada and by the Province of Ontario through the Ministry of Economic Development, Job Creation and Trade. The work of BCz is supported by Tsinghua University and the Thousand Young Talents Program.

\end{document}